\documentstyle[preprint,aps]{revtex}
\input epsf.tex
\def\beqa{\begin{eqnarray}}
\def\eeqa{\end{eqnarray}}
\def\beq{\begin{equation}}
\def\eeq{\end{equation}}
\def\beqan{\begin{eqnarray*}}
\def\eeqan{\end{eqnarray*}}

\def\half{\frac{1}{2}}

\def\ln{{\rm ~ln}}

\def\hm{h^{-1}{\rm~Mpc}}

\def\ddemu{_{;\mu}}  \def\udemu{^{;\mu}}

\def\lam{\lambda}  
\def\Lam{\Lambda}
\def\eps{\varepsilon}

\def\Gam{\Gamma}

\def\ome{\omega}

\newcommand{\mincir}{\raise -2.truept\hbox{\rlap{\hbox{$\sim$}}\raise5.truept
\hbox{$<$}\ }}
\newcommand{\magcir}{\raise -2.truept\hbox{\rlap{\hbox{$\sim$}}\raise5.truept
\hbox{$>$}\ }}
\newcommand{\minmag}{\raise-2.truept\hbox{\rlap{\hbox{$<$}}\raise 6.truept\hbox
{$>$}\ }}

\def\etal{{\it et al.}\ }

\def\rapj{ Ap. J. }

\begin{document}
\draft
%\title{{\large {\bf Classical, quantum and gravitational corrections to bubble
%nucleation in first order inflation}}}
\title{{\large {\bf 
Reconstruction of the bubble nucleating potential}}}
\author{Luca Amendola$^{1}$, Carlo Baccigalupi$^{1,2}$, \\
Rostislav Konoplich$^3$, Franco
Occhionero$^1$ \& Sergei Rubin$^3$}

\address{$^1$Osservatorio Astronomico di Roma,
Viale del Parco Mellini, 84, 00136 Roma, Italy\\
$^{2}$ Dipartimento di Fisica, Universit\`a di Ferrara,
Via Savonarola 9, 44100 Ferrara, Italy\\
$^3$Moscow Engineering Physics Institute, Kashirskoe sh, 31, 115409 Moscow,
Russia}
\maketitle

%\date{\today}
\maketitle
 
\baselineskip 14pt
\vspace{2.cm} 
\begin{abstract}

We calculate analytically the bubble nucleation rate in a model
of first order inflation which is able to produce large scale
 structure. The computation includes the first-order
departure from the thin-wall limit, the explicit
derivation of the pre-exponential factor, and the gravitational
correction.
The resulting bubble spectrum
 is then compared with constraints from the
large scale structure and the microwave background.
We show that there are models which pass all the constraints and
produce bubble-like perturbations
of interesting size. Furthermore, we show that it is in
principle possible to  reconstruct completely the
inflationary two-field potential from observations.

\end{abstract}
 
%\vfill	
%\centerline{To be published on Phys. Rev. D}
\vspace{.3cm}
 
%\newpage

	\section{\normalsize\bf Introduction.}

One of the most interesting ideas introduced in inflationary cosmology
in recent years is the possibility of performing a
phase transition {\it during} inflation.
In such scenarios, two fields act on stage: one, say $\ome$, slow
rolls, driving enough inflation to solve the standard problems;
the second field, say $\psi$, tunnels from a false vacuum state to an
energetically favoured true vacuum state, producing
{\it bubbles} of the new phase embedded in the old one. 
Both processes are governed by a two-field
potential $U(\ome,\psi)$.
To avoid the graceful exit
problem, the true vacuum state has to allow for a period of inflation
on its own.
We can then speak of a true vacuum {\it channel} over which the
bubbles slow roll until inflation ends, and reheating takes over.
Depending on the potential, three classes of first-order inflation
models
have been proposed so far.
The first is the classical extended inflation
\cite{LS,LSB,AF}: the bubbles
are produced in a copious quantity, so that they eventually
fill the space and complete the transition. To avoid
too large
distortions on the CMB, this scenario must produce very small
bubbles\cite{TWW,LW},
 so that  they are rapidly thermalized after inflation.
No trace of the bubbles is left in our Universe, and from this
point of view such scenarios do not lead to new predictions over
inflation without bubble production.
The second class is the $\Omega<1$ inflation\cite{BGT,LIN,ABO}:
 here the transition
is never completed, so that each bubble resembles an open Universe
to inside observers. Therefore,  if the bubbles inflate for
less than the canonical $N_T=60$ or so $e$-foldings, 
they will approach an $\Omega<1$
Universe. Here the effect of the nucleation process is 
observable, although it is indistinguishable
from a slow roll inflation just shorter than
$N_T$ $e$-foldings and with no first-order
phase transition at all.
Finally,  in \cite{OA} a third class of models has been proposed,
following an early suggestion of La \cite{LA}.
In such models, the phase transition is completed {\it before}
the end of inflation. Then, it has been shown
 that the primordial bubbles can be
large enough to drive structure formation, and still be
below the CMB level of detection\cite{AO,AB,OBA,BAO}.
 In such a scenario,
the present large scale structure is a direct outcome of
the first order transition, which is therefore observable and testable.

Several deep redshift surveys detected large voids in the galaxy
distribution\cite{cfa,lascamp,piran}, 
although it is still not clear if  they are
really empty of matter or just lack luminous galaxies \cite{lya}.
Standard models of galaxy  formation can barely accounts for these
structure, and do so only at the price of adjusting the parameters
to get very large scale power (see e.g. Ref. \cite{boh}). 
Therefore, just as we associate matter clumps to primordial fluctuations,
it appears 
worth trying to associate the present voids to primordial bubble-like
fluctuations, produced during a first-order phase transition.
Within different contexts, the idea of the voids as 
separate dynamical entities has been investigated several times
in earlier literature \cite{OC}.

A crucial aspect of bubble inflation is the calculation
of the bubble spectrum $n_B(L)$, defined as the number of 
bubbles per horizon with comoving size larger than $L$.
In \cite{OA}
we calculated $n_B$ in a specific model, built on fourth order gravity
\cite{S},
which we found to possess the requested features. 
We found that $n_B(L)$ can be approximated by a power law,
\beq
n_B=(L_{m}/L)^ p\,,
\eeq
 and that $L_{m}$ can be
as large as the observed voids in the Universe.

The central quantity needed to  evaluate $n_B$ is the  nucleation rate
in the semiclassical limit
\cite{Coleman}
\beq\label{tunnrate}
\Gamma={\cal M}^4 \exp (-B)
\eeq
 where ${\cal M}$ is a constant with a dimension
 of mass, and $B$ is
the  Euclidean least action minus the action
for the external deSitter spacetime solution.
The calculation of $\Gamma$ in \cite{OA} (and in
most other papers on the topic)
was based on the thin-wall limit, on neglecting the gravitational
correction, and on a dimensional argument for evaluating ${\cal M}$.
In this paper we remove, at least partially,
 all three approximations. 
 This calculation allows us to reconstruct completely
 the inflationary two-field potential from the
 determination of four observable quantities:
 the slope of the bubble spectrum, its amplitude, the density
 contrast inside the  bubbles, and the amplitude of the
 ordinary slow-rolling fluctuations. We remark that
 only in a model, like ours,
 in which the bubbles are directly observable, it is
 possible to reconstruct the tunneling sector of the primordial
 potential.
 
 The scheme of the paper is as follows.
In Sec. 2 we present the class of potentials we are going to investigate.
In Sec. 3 we give the basic formulas. In Sec. 4,5, and 6 we 
go through the detailed calculation of $\Gamma$, taking into
account deviations from the thin wall limit, inserting
the gravitational correction and calculating explicitely the
factor ${\cal M}$.
%We anticipate that the result is that $n_B$ can still
%be approximated as a power law, with the
%observable quantities $p$ and $L_{m}$
%function of the potential parameters. 
In Sec. 7
we introduce our model of first-order inflation in a fourth-order
gravity theory \cite{S} and calculate the time-dependent nucleation rate. In
Sec. 8 we write down the bubble spectrum
and in Sec. 9 finally we compare it with constraints from  CMB and 
large scale structure. In the last section we draw our conclusions.

\section{The model}

We consider the scalar field theory described by the action
(hereinafter, $\hbar=c=1$)
\beq\label{norms}
S=\int d^4x\sqrt{-g} \left\{ -{{\cal R}\over 16\pi G}
+\half\psi\ddemu\psi\udemu-U(\psi)
\right\}
\eeq
where $g$ is the metric determinant and ${\cal R}$ the curvature
scalar.
The potential is a generic quartic function 
with non-degenerate minima which allows for tunneling. We can write it
very generally in the form
\beq\label{general}
U(\psi)=\Lam+V_1(\psi)+V_2(\psi)
\eeq
where $\Lam$ is a cosmological constant, $V_1$ is a quartic
with two equal-energy minima and $V_2$ is a symmetry-breaking potential
which brings the energy of one minimum, the false vacuum (subscript F),
to a  value  larger than the other, the true vacuum (subscript T).
In the two minima, the Einstein equation reduces simply
to 
\beq
H^2=8\pi G U/3
\eeq
We will denote with $H_T,U_T$ the Hubble constant and the potential energy
 of the true vacuum, and with $H_F,U_F$ the same quantities
for the false vacuum.
We wish to calculate the tunneling rate (\ref{tunnrate})
where $B=S_E(\psi)-S_E(\psi_{F})$, and $S_E(\psi)$
 is the  Euclidean least action, i.e. the Action of the
 ``bounce'' solution.
We perform the computation  in the
thin-wall limit (actually we go to a
post-thin-wall limit calculation),
 according to which the $O(4)$ bubbles nucleated
have 4-radius $R\gg \Delta$, where $\Delta$ is the wall thickness.
Further, we include the gravitational term in the action: as we
will show, this term
is important when the parameter $g=R H_T$ (not to be
confused with the metric determinant) is much larger than unity.
This limit amounts in fact to a bubble approaching  the
space curvature radius $1/H$. 
It is convenient to write
the potential (\ref{general}) in the form
\beq\label{potphys}
U(\psi)={3g^2\over 8\pi G R^2} +{1\over 2\Delta^2\psi_0^2}
(\psi^2-\psi_0^2)^2+{1\over R\Delta}(\psi+\psi_0)^2
\eeq
Then, the true vacuum state is $\psi=-\psi_0$, and the false vacuum
is $\psi=\psi_0$. 
The potential (\ref{potphys}) is therefore
defined by four physical parameters: $g,R,\Delta,\psi_0$.

\section{Thin wall limit}

The Euclidean action of the scalar theory (\ref{norms}) is
\beq\label{grea}
S_E=\int d^4x\sqrt{-g} \left\{ -{{\cal R}\over 16\pi G}
+\half\psi\ddemu\psi\udemu+U(\psi)
\right\}
\eeq
In the Euclidean metric for a O(4) space
$ds^2=dr^2+a^2( r) d\Omega_3^2$ one has
${\cal R}=-6(aa''+a'^2-1)/a^2$ and
\beq\label{secalc}
S_E=2\pi^2\int d r\left[{3\over 8\pi G}(a^2 a''+a a'^2-a)
+a^3(\half\psi'^2+U)\right] 
=-{3\pi\over 2 G}\int d r \,\,a( 1-a^2 H^2)
\eeq
where the prime denotes derivation with respect to the 4-radius $r$.
The Euclidean Klein-Gordon
 equation for $\psi$ is
\beq\label{eepsi}
\psi''+3{a'\over a} \psi'=dU/d\psi
\eeq
and the Euclidean Friedmann equation is
\beq\label{eufri}
a'^2=1+{8\pi G\over 3} a^2(\half \psi'^2-U)
\eeq
In the zero-gravity limit, $G\to 0$, the latter equation gives
$a'={\rm const}$, so that (\ref{eepsi}) reduces to
\beq\label{euzero}
\psi''+{3\over r}\psi'=dU/d\psi
\eeq
In the thin-wall limit, in which $R\gg \Delta$, one can
assume that the second term in (\ref{euzero}) can be neglected,
and that $dU/d\psi=2\psi(\psi^2/\psi_0^2-1)/\Delta^2$.
The solution which interpolates
between false and true vacuum is then
\beq
\psi^{(0)}=\psi_0 \tanh \left({r-R_w\over \Delta}\right)
\eeq
where $R_w$ is an integration constant that will be determined later.
To integrate the action over the bounce solution, we consider
that outside the bubble,
i.e. in the false vacuum,  $\psi=\psi_0$, so that $B_{ext}=S_E(\psi_0)
-S_E(\psi_0)=0$.
On the wall,  at distance $R_w$, 
we have 
\beq\label{bwall}
B_{wall}
=2\pi^2  R_w^3 S_1
\eeq
where
\beq\label{s1}
S_1=\int_{-\psi_0}^{\psi_0} d\psi
\left[2(U(\psi)-U_F\right]^{1/2}={4\psi_0^2\over 3\Delta}
\eeq
Finally, inside the bubble, $\psi=-\psi_0$, and since
(from (\ref{eufri}), and neglecting $\psi'$)
$da=d r \left[1-a^2 H^2_T\right]^{1/2}$
we have
\beq
B_{int}=-{3\pi\over 2G}\int_0^{ R_w} a da
\left[(1-a^2 H^2_T)^{1/2}-
(1-a^2 H^2_F)^{1/2}\right]
\eeq
The general expression is therefore 
\beq\label{genb}
B(R_w)=
2\pi^2 R_w^3 S_1+{\pi\over 2G}\left\{
H_T^{-2}\left[(1-R_w^2 H_T^2)^{3/2}-1\right]
-H_F^{-2}\left[(1-R_w^2 H_F^2)^{3/2}-1\right]\right\}
\eeq
Let us note that $R=3S_1/\eps$, where 
$\eps=U_F-U_T=4\psi_0^2/R\Delta$. Then we see that $B(R_w)$ is minimized by
\cite{Parke}
\beq\label{genrw}
R_w= R \left(1+g^2+4\pi G R S_1+12\pi^2 G^2 R^2 S_1^2\right)^{-1/2}
\eeq
Then, for $G\to 0$, which implies 
$g=RH_T \to 0$, one has
$R_w=R$. Notice that
the parameter $R_w H_T$ which appears in (\ref{genb}) equals
$g$, since $H_T^2=g^2/R^2$ for the potential (\ref{potphys}).
This shows explicitely the role played by the constants $R$
and $g$.
Finally, we obtain the usual
(zero-gravity, thin-wall) result\cite{Coleman}
\beq\label{ea}
B_0=27\pi^2{S_1^4\over 2 \eps^3}=
{2\pi^2\over 3} {R^3\psi_o^2\over \Delta}
\eeq

\section{The gravitational correction}

Now to the general case, $G\not = 0$. 
To the purpose of this paper, we simplify our problem
by putting $(H_F^2-H_T^2)/H_T^2\ll 1$, i.e. assuming that
the vacuum energy difference is much smaller than the true vacuum energy.
Our results will be consistent with
this approximation.
This is equivalent to neglecting the last two terms in 
parentheses in Eq. (\ref{genrw}). Then we have 
that $B(R_w)$ is minimized by
a bubble radius 
$R_w =R/(1+g^2)^{1/2}$. The bounce action is
$B=B_0 f(g)$
where $B_0$ is the no-gravity action (\ref{ea}) and where
\beq
%\label{bgrav}
f(g)=4(1+g^2)^{-3/2}\left\{1+g^{-4}[2+3g^2-2(1+g^2)^{3/2}]\right\}
\eeq
For $g\to 0$, $f(g)\to 1$, as expected. To the lowest non-trivial
 order in $g$,
\beq\label{bgrav}
B=B_0(1-g^2)
\eeq
Gravitational effects, then, increase the nucleation rate 
$\Gamma$  \cite{Coleman}.

\section{\normalsize
	\bf Post-thin-wall correction}

Let us come back to
the Euclidean spherically symmetric equation of motion 
in absence of gravity, Eq. (\ref{euzero}).
%\beq
%\psi''+{3\over r}\psi'=dU/d\psi
%\eeq
While the trivial true-vacuum solution 
  $\psi_{-}=-\psi_0$ holds also when the symmetry-breaking term
  in the potential is considered, the false vacuum solution is
  slightly displaced:
\beq
\psi_{+}=\psi_0\left (1-{\Delta\over R}-{\Delta^2\over R^2}-....\right)
\eeq
It is useful now to expand the equation of motion around
$r=R$. Introducing the variable $z=(r-R)/\Delta$ we obtain
\beq\label{eqz}
\psi''+{3\Delta\over R}\psi'(1-{z\Delta\over R}+
{z\Delta^2\over R^2}+...)={2\psi\over
\Delta^2}(\psi^2/\psi_0^2-1)+{2\over R\Delta}(\psi-\psi_0)
\eeq
where now the prime denotes derivation with respect to $z$. 
We search solutions of Eq. (\ref{eqz}) to the first order
in $\Delta/R$:
\beq
\psi=\psi^{(0)}+\psi^{(1)}(\Delta/2R)+...
\eeq
We know already that to the zero-th order
$
\psi^{(0)}=\psi_0\tanh z
$.
Subtracting the zero-th solution from Eq. (\ref{eqz}) we
get, for $i\ge 1$
\beq
\psi''^{(i)}+\psi^{(i)}\left[-4+{6\over \cosh^2 z}\right]=f_i(z)
\eeq
where $f_i(z)$ is defined order by order from Eq. (\ref{eqz})
and depends only on $z$ and not on $\psi^{(i)}$.
At first order, then, the solution is 
\beq\label{cla}
\psi=\psi_0\left\{\tanh z+\left({\Delta \over 2R}\right)
\left[ -{\frac z{\cosh ^2z}}%
-\tanh z-1\right] -2\left(\frac \Delta {2R}
\right)^2\left[ \text{tanh}z+1\right]\right\}
\eeq
It can be noticed, from Fig. 1, that the post-thin-wall correction
moves the large-$z$ value of $\psi$ from $\psi_0$ to $\psi_+$.
In Eq.(\ref{cla}) we included the part of the correction at second order
which
also contributes to the leading classical correction to the action after
integration over $z$.

The Euclidean action, finally, is calculated as
\beq\label{bpost}
B=S(\psi)-S(\psi_0)=
B_0(1-9\Delta/2R)
\eeq
The term $9\Delta/2R$ is therefore the post-thin-wall
correction term, which again increases the tunneling rate.

\section{The pre-exponential factor}

In semiclassical approximation in four dimensions the rate of false vacuum
decay is given by
Coleman's formula \cite{Coleman}
\begin{equation}
\label{colm}\Gamma =\left( \frac B{2\pi }\right) ^2\left| \frac{\det
^{\left( ^4\right) }\left( -\partial ^2+U^{\prime \prime }\left( \psi
\right) \right) }{\det \left( -\partial ^2+U^{\prime \prime }\left( \psi
_{+}\right) \right) }\right| ^{-1/2}e^{-B},
\end{equation}
where the functional determinant is computed with the four zero eigenvalues
omitted, $\psi $ is the classical solution of Eq. (\ref{eqz}), $\psi _{+}$ is
the trivial false vacuum solution. It was shown in \cite{RVKon} that in the
thin wall limit the functional determinant can be expanded in power series
of $2R/\Delta $ as

\begin{equation}
\label{Rmu}{\rm ~ln}\frac{\det ^{\left( 5\right) }D}{\det ^{\left( 5\right)
}D_0}=C_l{\rm ~ln(}2R/\Delta )+C_3\left( 2R/\Delta \right) ^3+C_2\left(
2R/\Delta \right) ^2+...
\end{equation}
The parts of this expression, which behave as powers of $2R/\Delta $ have no
universal significance, for they are adjusted or even completely concealed
by ultraviolet renormalisations and we will include them in overall
multiplicative factor ${\cal M}^{\prime }{\cal .}$ The logarithmic part, on
the other hand, is not expected to be affected by short distance structure.
The universal infrared logarithm ${\rm ~ln}(2R/\Delta) $ is associated with
the geometrical features of the bubble and does not depend on the
specific form
of the double well potential. Since the logarithmic term appears due to 
the infrared
region we have to consider the low-lying levels of the operator of the
small fluctuations around the bounce solution

\begin{equation}
\label{fluct}D=-\partial ^2+U^{^{\prime \prime }}\left( \psi \right) .
\end{equation}
This operator is rotationally invariant and thus its eigenfunctions are, in
their angular dependence, 4-dimensional scalar spherical harmonics. In terms
of radial eigenfunctions the eigenfunction equation becomes

\begin{equation}
\label{eigen}\left[ -\frac{d^2}{dr^2}+\frac{l(l+2)+3/4}{r^2}+U^{^{\prime
\prime }}\left( \psi \right) \right] f_{nl}(r)=\lambda _{nl}f_{nl}(r)
\end{equation}
Expanding the centrifugal potential in power series around $r=R$ (it is
justified when we deal with the low lying states $l\ll 2R/\Delta $ since
bound states are localized at the wall of the bubble), keeping only the
leading term, and replacing $r$ with $z$,
 we can rewrite Eq. (\ref{eigen}) for $n=0$ as

\begin{equation}
\label{eigen2}\left[ -\frac{d^2}{dz^2}+\Delta^2 U^{^{\prime \prime
}}\left( \psi \right) -2E_0\right] f_{0l}=0,
\end{equation}
where
\begin{equation}
\label{E0}E_0=\frac 12\left[\Delta^2 \lambda_{0l}-4-\frac{4l(l+2)+3%
}{(2R/\Delta )^2}\right]
\end{equation}
does not depend on the quantum number $l$. Taking $E_0$ in the form of
expansion
\begin{equation}
\label{E02}E_0=2\left[ a+\frac b{2R/\Delta }+\frac c{(2R/\Delta )^2}\right]
\end{equation}
we obtain the spectrum of the eigenvalues in the following form
\begin{equation}
\label{lambda}\lambda _{0l}={4\over \Delta^2}\left[ a+\frac b{2R/\Delta }+\frac
c{(2R/\Delta )^2}+1+\frac{4l(l+2)+3}{4(2R/\Delta )^2}\right] .
\end{equation}
We know that in the spectrum of the eigenvalues there is one negative
eigenvalue $\lambda _{00}$ corresponding to uniform expansion of the bubble,
which is just vacuum instability. There is also the
4-fold degenerate eigenvalue
$\lambda _{10}=0$ describing translations of the bubble center in the
Euclidean space in four directions. Therefore, substituting $l=1$ into Eq.%
(\ref{lambda}) we obtain
\begin{equation}
\label{abc}a=-1,b=0,c=-15/4
\end{equation}
and find the lowest band of the eigenvalues
\begin{equation}
\label{lambda2}\lambda _{0l}=\frac{\left( l-1\right) \left( l+3\right) }{R^2}
\end{equation}
It is important to know that in deriving Eq. (\ref{lambda2}) for $\lambda _{0l}$
we assumed the existence of nontrivial classical solution of the equation of
motion which break a translational invariance (zero modes) and the thin wall
limit in order to justify the expansion in powers of $2R/\Delta .$ Therefore
this equation does not depend on specific form of the potential $U\left(
\psi \right) .$

Calculating the product of $\lambda _{0l}$ for $l\ll 2R/\Delta $ and keeping
only the
logarithmic term it is strightforward to show that  $%
c_l=-10$ (see Eq. \ref{Rmu}). Then it follows immediately from the general
formula (\ref{colm}) that in the thin wall limit, and
 taking into account the
quantum corrections, the rate of false vacuum decay is given by
\begin{equation}
\label{G2}\Gamma =\frac{{\cal M}^{\prime 4}B^2}{\left( 2R/\Delta
\right) ^4}e^{-B}
\end{equation}
independently of the specific form of the potential $U\left( \psi \right) .$
%Therefore classical (\ref{bpost}) and infrared quantum corrections lead 
%to a significant increase in the nucleation rate.

\section{A fourth-order gravity, first-order inflation model}

As we mentioned in the
Introduction, astrophysically interesting bubbles can be produced only
in a two field model, in which one has slow-rolling along
a field,  $\omega$, while the second field,  $\psi$, performs a 
phase transition in an orthogonal direction. 

We already introduced in \cite{OA} a specific model of
first-order inflation able to generate large bubbles
in fourth order gravity.
Once a conformal transformation is applied, the model is described
by the action (we put hereinafter also $G=1$ )
\begin{equation}\label{eua}
S=\int\sqrt{-g} d^4x \left[ -{{\cal R}\over 16\pi}
+(3/4\pi)\ome\ddemu\ome\udemu+
{1\over 2} e^{-2\omega}\psi_{;\mu}\psi^{;\mu}
+U(\psi,\omega)\right]\,,
\end{equation}
where the potential is
	\begin{equation}\label{conpot}
	U(\psi,\omega)
	=e^{-4\omega}\left[V(\psi)+{3M^2\over 32\pi}
	W(\psi)(1-e^{2\omega})^2  \right]\,,
	\end{equation} 	
	and where
\begin{equation}\label{ansatz}
	W(\psi)=1+{8\lambda\over \psi_0^4} \psi^2
        (	\psi-\psi_0)^2\,, \quad
		V(\psi)={1\over 2}m^{2}\psi^{2}\,.
	\end{equation}
	The slow roll inflation driven by $\omega$ takes place at $\omega\gg
	 1$,
	and is over when $\omega$ approaches zero.
	At large $\omega$, the potential $U$ is dominated by $W(\psi)$,
	and thus the false vacuum minimum at 
	$\psi\approx \psi_0$, for which $U_F=e^{-4\ome}[(3M^2/32\pi)
	(1 - e^{2\ome})^2+ V (\psi_0)]$
	is unstable with respect to tunneling toward
	the true vacuum
	$\psi=0$ (for which $U_T\approx
	3M^2/32\pi$). At small $\ome$, $U$ is dominated by
	$V(\psi)$, and both the true and the false vacua converge
	to the global zero-energy minimum at $\ome=\psi=0$,
	where inflation ends and  reheating takes place.
	The slow-roll solution in this model for $\ome\gg 1$ 
	can be written very
	conveniently as \cite{OA}
\begin{equation}\label{ef}
{4\over 3}N =e^{2\omega}\,.
\end{equation}
where $N$ is the number of $e$-foldings to the end of inflation.
In the same limit, $H=M/2$.
The potential (\ref{conpot}) for $\omega\gg 1$
takes the form (\ref{potphys})
by the substitution:
\begin{equation}
\label{psih}\psi \rightarrow \psi +\psi _0/2
\end{equation}
and therefore in the new notations we have:
\beqa
R&=&  (3/2\pi)^{1/2}
{M\lam^{1/2}\over m^2\psi_0} e^{4\ome}\\
\Delta &=& (8\pi/3)^{1/2}
{\psi_0\over M\lam^{1/2}}\\
g &=& MR/2
%(3/8\pi)^{1/2} {M^2\lam^{1/2}\over
%m^2\psi_0}e^{4\ome}
\eeqa
%\beqa
%R&=&  (6/\pi)^{1/2} {M\lam^{1/2}\over m^2\psi_0} e^{4\ome}\\
%\Delta &=& (32\pi/3)^{1/2} {\psi_0\over M\lam^{1/2}}\\
%g &=& (3/2\pi)^{1/2} {M^2\lam^{1/2}\over m^2\psi_0}
%e^{4\ome}
%\eeqa
Notice that now $R$ depends on the $e$-folding time as $R\sim N^2$. This is
where the scale-dependence of the bubble spectrum arises from, as we 
show below.

The Euclidean action is	
\begin{equation}
%\label{eua}
S_E=\int\sqrt{-g} d^4x \left(-{{\cal R}\over 16\pi}
+{1\over 2} e^{-2\omega}\psi_{;\mu}\psi^{;\mu}
+U(\psi,\omega)\right)
\end{equation}
(neglecting the kinetic energy of $\omega$).
To obtain a canonical kinetic term, we rescale the
coordinates \cite{kolb} as
$~ x^{\mu}=e^{-\omega} \hat x^{\mu}\,,~$
so that,  in the new coordinates
\begin{equation}\label{newact}
S_E=e^{-4\omega}\int\sqrt{-g} d^4\hat x \left(-{{\cal R}\over 16\pi}
+{1\over 2} \psi_{;\mu}\psi^{;\mu}
+U\right)=e^{-4\omega} \hat S_E(\psi)\,,
\end{equation}
where finally $\hat S_E$ is canonical. 
Combining  Eq. (\ref{newact}),
Eq. (\ref{bpost}) and Eq. (\ref{bgrav}), we obtain
\beq
B=e^{-4\ome} {\pi^2\over 6} {R^3\psi_0^2\over\Delta}
\left(1-{9\Delta\over 2R}\right)(1-g^2)
\eeq
(there is an extra factor of 1/4 because now the vacuum states are at
$0$ and $\psi_0$ instead of at $\pm\psi_0$).
In terms of $N$ this is
\beq\label{bn}
B={N^4\over N_1^4}\left[1-\left({N_2\over N}
\right)^2\right]\left[1-\left({N\over N_3}
\right)^4\right]
\eeq
where
\beqa
N_1^2 &=&  {3^{3/2}\over 4}{\psi_0 m^3\over M^2\lam}%
\\
N_2^2 &=& {27\pi\over 8}{\psi_0^2 m^2\over M^2\lam}\\
N_3^2 &=&
\left({27\pi \over 32}\right)^{1/2} {\psi_0  m^2\over M^2\lam^{1/2}}
%N_1^2 &=&  (\sqrt{27}/16){\psi_0 m^3\over M^2\lam}\\
%N_2^2 &=& (27\pi/8){\psi_0^2 m^2\over M^2\lam}\\
%N_3^2 &=& (27\pi/128)^{1/2} {\psi_0 m^2\over M^2\lam^{1/2}}
%={m\over 3M} N_2
\eeqa
The two approximations we adopted are valid for
$N>N_2$ (thin-wall), and for $N<N_3$ (gravitational correction). 
It is useful now to introduce a fourth $e$-folding epoch, $N_0$,
defined as the epoch at which the crucial quantity $Q=4\pi\Gamma/9 H^4$
equals unity. This can be seen as the epoch at which  one bubble for 
 horizon volume for Hubble time (or 4-horizon) is nucleated,
  that is, the bubbles
 are saturating the false vacuum space. The bulk of the
 nucleation occurs therefore just before $N_0$, ans spans only
 few $e$-folding times. We can therefore approximate $Q(N)$ as
\beq
Q={4\pi \Gam \over 9H^4}=\exp \left\{
{(N_0^4-N^4)\over N_1^4}\left[1-\left({N_2\over N_0}
\right)^2\right]\left[1-\left({N_0\over N_3}
\right)^4\right]\right\}
\eeq
where 
\beq\label{n0}
N_0 ^4= \frac{N_1 ^4}{[1-(N_2 /N_0)^2][1-(N_0/N_3 )^4]}
\ln \left({64 \pi {\cal M}^4\over 9 M^4}\right)
\eeq
From Eq. (\ref{G2}) we can write the factor ${\cal M}$ as
${\cal M}' B^{1/2}(\Delta/2R)$.

The scalar and gravitational zero-point fluctuations generated
during the $\omega$-driven slow-rolling are proportional to $M$ 
\cite{S83}.
From the current microwave
background measurements, we obtain $M\approx 5\cdot 10^{-6}$
(in Planck units),
a value that we will adopt in the  numerical results below.
However,
since in our model one should also consider
the contribution of the bubbles to the microwave background, this
constraint is actually only an upper limit on $M$.
Once $M$ is fixed, 
the three $e$-folding constants $N_{1-3}$ fully determine the
inflationary potential, through the three remaining constants
$m,\lam,\psi_0$. 
A successful sequence of epochs is the following:
 at $N_3$, the gravitational correction ceases to be
 important; 
 at $N_0$ the crucial parameter $Q$ 
 is of order unity, and the bubbles
 saturate the space; at $N_2$ the thin wall
 approximation breaks down; and finally at $N_1$
  the bounce action $B$ equals unity, and thus
 the semi-classical approximation is no longer reliable. 
 For traces of the phase transition to be visible
 today, we must also impose $N_0<N_T$, being $N_T$
 the observable last 60 $e$-foldings of the inflation.
 Then, the full set of constraints is
 \beq\label{fullset}
 N_3>N_T>N_0>N_1,N_2\,.
 \eeq

\section{The bubble spectrum}

We can now calculate explicitely the bubble spectrum in our model.
  The number of bubbles nucleated in the interval $dt$ is \cite{GW}
\begin{equation}\label{ourspe}
{dn_B\over  dt}=\Gamma a^3 V_{in} \exp (-I)\,,
\quad I\equiv \left\{-{4\pi \over 3} 
\int_{0}^{t} dt'\Gamma(t') a^3(t')
\left(\int_{t'}^{t} {d\tau \over a(\tau)}  \right)^3\right\} \,,
\end{equation}
	where $V_{in}$ is the horizon volume at $N=N_T$, 
	$V_{in}=4 \pi /3 H_{in}^3$, and 
	where the exponential factor accounts for the fraction of
	space which remains in the false vacuum.
To get a manageable expression,
 we first change variable in Eq. (\ref{ourspe}) from
the nucleation epoch $t$ to the scale $L$ in
horizon-crossing at $t$, by use of the relation $dL/dt\approx
-H_{in} L_h/a$ valid during slow roll. 
This gives
\beq
{dn_B\over dL}=-3L_h^3L^{-4} Q(N) e^{-I}\,.
\eeq
We approximate $Q(N)$ around $N=N_0$ as $Q\approx \exp [s(N_0-N)]$,
where
\def\lton{\left(}
\def\rton{\right)}
\def\lqua{\left[}
\def\rqua{\right]}
\beq\label{snnn}
s=4{N_0^3\over N_1^4}\lqua 1-\lton{N_2\over N_0}\rton^2\rqua
\lqua 1-\lton{N_0\over N_3}\rton^4\rqua
\eeq
We make  use of the relation between the $e$-folding time $N$
 and the bubble comoving size $L$  
\begin{equation}\label{hln}
HL(N)=H_{in}L_h \exp (N-N_T)\,.
\end{equation}
It follows $Q=e^{-s\Delta N} (L_h/L)^s$,
where $\Delta N=N_T-N_0$ corresponds to the duration of
the transition.
For $I\ll 1$, i.e. far from the end of the transition, we obtain
\beq
{dn_B\over dL}=A L^{-4-s}
\eeq
and
\beq\label{ourlaw}
n_B=(L_m/L)^{p}\,,\quad p=3+s
\eeq
where
\beq\label{newlm}
L_{m}=L_h e^{(3-p) \Delta N/p} (3/p)^{1/p}
\eeq
In the thin-wall, zero-gravity limit, $s=4N_0^3/N_1^4$ \cite{OA}.
When $I$ approaches unity, the nucleation process reaches a peak, and then declines 
rapidly, due to the fast decrease in the false vacuum space available. 
In \cite{OA}
we showed that the peak occurs, as expected,
 just after $N_0$, which we therefore consider the end of the nucleation.

As we discussed in Ref. \cite{OA}, Eq. (\ref{newlm}) differs
from the analogous quantity in extended inflation \cite{LS}\cite{LW} \cite{LA}
because here we have $\Delta N$ instead of $N_T\approx 60$.
We show below,
by a suitable choice of $\Delta N$, i.e.  of the duration of the transition, 
that we may have a bubble
spectrum $n_B$ which generates astrophysically large bubbles
and escapes the CMB bounds, contrary to what occurs in the
other models  of first-order  inflation.

\section{\normalsize
	\bf Results and Discussion.}

We wish now to compare the bubble spectrum with the
constraints from large-scale structure and the cosmic
microwave background.
It is not the aim
of this work to determine the best model parameters, also because
the statistics on observed voids is still at a very schematic
level. We will focus instead on
showing that there is a region of parametric space
which gives astrophysically large bubbles, 
that can contribute significantly, if not exclusively,
to structure formation,
while passing
the CMB constraints.

When a bubble nucleates, its density contrast can be estimated
(in the thin-wall limit),  as
\beq
\delta\equiv |\delta\rho/\rho|={U_F-U_T\over U_F}=[(N/N_4)^2+1]^{-1}
\eeq
where 
\beq
N_4^2=3\pi {\psi_0^2 m^2\over M^2}
\eeq
Since the bubbles cross out the horizon soon after their
nucleation, the same density contrast will be
found at reenter.
The bubble density contrast  increases to unity for
$N\ll N_4$, i.e. toward the end of
inflation. One can also write $N_4\approx 1.5 M N_3^4/N_1^2 $.
%Here we consider small values of $\delta$.
For instance, taking acceptable values as $N_3=90$, $N_1=20$ and
$M=5\cdot 10^{-6}$, one obtains $\delta\approx 6\cdot 10^{-4}$ at
$N\approx N_0\approx 50$. Taking $N_3\to\infty$ one would obtain completely
void bubbles, $\delta\to 1$.
Here we will consider instead only small values
of $\delta$ \cite{OBA}.
From 
Eq. (\ref{hln}) and $N\gg N_4$, we have the density spectrum at reenter
\beq
\delta(L)=\left({N_4\over N_T}\right)^2\left[1-{2\over N_T}\log (L/L_h)\right]
\eeq
Since $\delta\ll 1$, we do not consider (contrary to 
Ref. \cite{OA,AO,BAO}) the overcomoving
growth that takes place only if and when, in its late history, the
density becomes non-linear.

Let us discuss first the CMB constraints.
A bubble of radius $L$ at decoupling produces
a Sachs-Wolfe distortion on the microwave temperature of
$\Delta T/T\sim \delta(L) L^2/L_d^2$, if $L_d$ denotes the horizon scale
at decoupling. 
In reality, bubbles which reenter before decoupling
have time to deepen; for the range of scales
we are interested in, however, this is a minor effect and we neglect it.
In a pixel corresponding to a size of
$L_p>L$ at decoupling, a further  factor of $L^2/L_p^2$ smears the signal
\cite{LW}. 
There are two main CMB constraints arising from observational
upper bounds to
such Sachs-Wolfe effect. Full-sky, low-resolution surveys like COBE can detect
rare big bubbles as hot spots. On the other hand,
a large number of small bubbles can be detected as Poissonian
fluctuations in high resolution, small
coverage experiments with antenna beam below $1^0$. 
Here we restrict our attention to the first kind of
observational effect, as the second one depends on the lesser
known physics of the small bubbles nucleated near the
end of the first order transition. At any rate, the small bubble
constraint is expected to be less severe than the hot-spot test
for bubble spectrum slopes $p\le 10$ \cite{AO}.
In \cite{OBA,BAO} we also analyzed the constraints on bubble models
from
Doppler effects on the last scattering surface, and found them
to be generally smaller  than the Sachs-Wolfe ones.

Assuming a power-law spectrum like (\ref{ourspe}) the
constraint can be put in form of restrictions on the two
parameters $p$ and $L_{m}$ for large $p$
\cite{AO}. Let $L_v$ denote the smallest bubble at decoupling that
can give an observable $3\sigma$ signal ($\Delta T/T\approx
5\cdot 10^{-5}$) in a COBE pixel. We simply require that
there are fewer than one bubble larger than $L_v$
intersecting the last scattering surface.
Then, the hot-spot constraint amounts to\cite{LW,OA}
\begin{equation}\label{both1}
L_{m}< 
L_v\left[{(p-1)L_h\over p L_v}\right]^{1/p}\,.
\end{equation}
$L_v$ is calculated as
\beq
{\Delta T/T}\approx {\delta(L_v)\over 3} {L_v^4\over L_d^2 L_p^2}
\eeq
We will take $L_d=190 \hm$, and $L_p=300 \hm$. The latter value corresponds to
a beam angular opening of $3^0$, which is
roughly  the COBE beam opening.

Next we impose to our bubble spectrum the
qualitative requirement that it be able
to produce a significant  large scale structure.
We realize this condition in a very simplified way: we just
find the parametric region for which bubbles larger than
3$\hm$ fill the space 
for more than 50\% \cite{OA}, a constraint
which can be compared, for instance, with the
observed voids in the SSRS2 database \cite{piran}. Notwithstanding its
simplicity, this minimal condition 
puts a strong constraint on the parameter space.
If we had indications on the real
matter content of large voids, we could 
put a further direct constraint on $\delta$. For instance,
if the voids of $20\hm$ (which are therefore nucleated
5-6 $e$-foldings after $N_T$)
that are currently detected on the
large scale surveys (see, e.g., \cite{piran} )
are completely empty of matter, they should have
$\delta\ge 10^{-3}$ at decoupling, in order for them
to be cleared of matter by linear growth at the present.
If some matter, maybe in the form of Ly$\alpha$ clouds \cite{lya}, is found
inside the voids, then one should impose on the contrary
$\delta < 10^{-3}$. Here, to fix the ideas,  we put $\delta=10^{-3}$
on scales of $\approx 30 \hm$. In any case,
the results are not very sensitive
 to $\delta$.
 
The main results are contained in Fig. 2.
On the plane $(p, L_m)$, we display as a shaded
area the parametric 
region of cosmological interest, i.e. the models  which
 satisfy
the CMB constraint (\ref{both1}),
and fill the space as requested.
The CMB constraint is plotted  for $N_3=90$ and 
$\delta=10^{-3}$, which, along with  $M=5\cdot 10^{-6}$, implies
 $N_1\approx 17$. 
 These values satisfy all the
conditions for a successful inflation. 
Then we plot two curves
 $L_m(p)$ given
in Eq.  (\ref{ourlaw}), for $N_0=53,55$. It can be seen
that these curves  cross
the acceptable region; our model is therefore
capable of producing
 bubble spectra  which pass the CMB tests and have
interesting large scale features.
Let us consider one pair of parameters for
reference: $p=9$, $L_m=100 \hm$, which lies on the curve
$N_0=54$, and  is inside the
acceptable region. Then, from
(\ref{snnn}), and putting
$\delta=0.001$ and $M=5\cdot 10^{-6}$,
 we have that the full set of $e$-folding parameters is
\beq
N_0=54\,, N_1=17\,, N_2=16\,, N_3=90
\eeq
which verifies (\ref{fullset});
 we have also  $N_4=1.7$.
Clearly, this is only one of an infinite set of
parameters that fulfills all the conditions; however, the values are
in general close  to those quoted. 
We may also note that with such a bubble spectrum, $p=9$ and $L_m=100\hm$,
one can estimate $(R_s/L_h)^3
n_B(>20\hm)\approx 70$
 voids larger than 20 $\hm$ by
radius in a spherical survey of depth $R_s=200 \hm$;
this value,
which of course
is extremely approximate, is indicative that many, if not all, of the 
present voids can be explained by a first-order phase transition.
Finally, the potential parameters,
other than $M$,
are (all in Planck units)
\beq
\lam=4\cdot 10^{-3},m=3.3 \cdot 10^{-3},\psi_0=1.8\cdot 10^{-3}
\eeq
Thus, as anticipated in the Introduction,
the observations of the two bubble spectrum
parameters, $L_m$ and $p$, 
of the matter content inside bubbles (so to fix $\delta$), and
of the ordinary fluctuations driven by $\omega$ on the microwave
background, which give $M$,
are in principle all what we need
to completely reconstruct the primordial first-order potential.
This would be impossible in the other models
of first-order inflation, in which bubbles do not
leave observable traces.

\section{\normalsize\bf Conclusions}

Large voids are ubiquitous in the Universe. They occupy
most of the observable volume, and are probably
the dominant contribution
to the large scale power spectrum \cite{lascamp}. The
standard possibility is that they simply derive from the
primordial underdensities which are expected in the ordinary Gaussian
models of structure formation. However, this scenario
would be put in jeopardy if larger and larger voids continue
to be discovered \cite{cohen}. Moreover, if the voids are actually
filled with unclustered, or mildly clustered, matter, it would
be impossible to explain their roughly spherical shape: an underdensity,
in fact, becomes more and more spherical only as it becomes
non-linear \cite{icke}. It is therefore worthy to investigate
alternatives. We proposed here, basing on earlier work \cite{AO}\cite{OA},
a first-order inflationary model which produces primordial non-empty
bubbles, along with ordinary slow-rolling fluctuations. We
calculated in detail the nucleation rate, including
classical, quantum and gravitational corrections, and showed that
the model gives a strong contribution to
large scale structure, while passing the microwave constraints.
The determination of four observable quantities 
 completely fixes the primordial potential, including the tunneling
sector, which, on the contrary, is unobservable in the other
models of first-order inflation.

\vspace{.5cm}
%\centerline{\bf Acknowledgments}

 \newpage
 \centerline{\bf FIGURE CAPTION}

 \begin{itemize}
 \item {\bf Fig. 1}
 
 The bounce solution, interpolating between false vacuum (FV) and
 true vacuum (TV). The solid line corresponds to $\Delta/R=0$,
 the dotted line to $\Delta/R=0.1$ and the dashed line to $\Delta/R=0.3$.

 \item {\bf Fig. 2}
 
 Parametric region of the model. The
 shaded area is the region which passes
 the constraint from the microwave background isotropy (solid
 line labelled CMB)
 and  the constraint from large scale structure (dashed line
 labelled LSS). The two dotted lines give $L_m=L_m(p)$
 in our model for the two values of $N_0$
 shown. The other parameters adopted
 are indicated in the box.
 \end{itemize}

 \newpage
 \epsfysize 6in
 \epsfbox{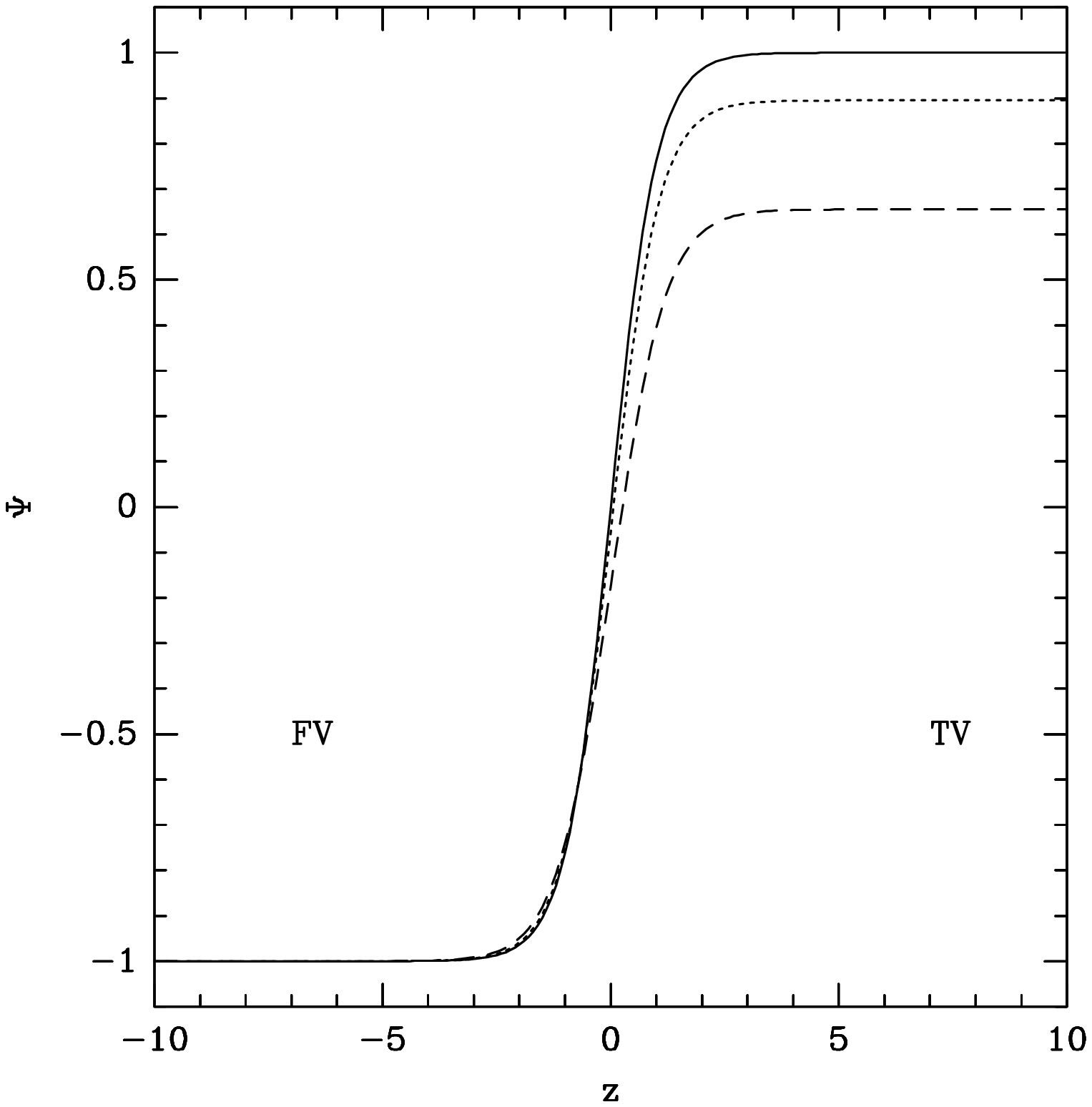}
 
  \newpage
 \epsfysize 6in
 \epsfbox{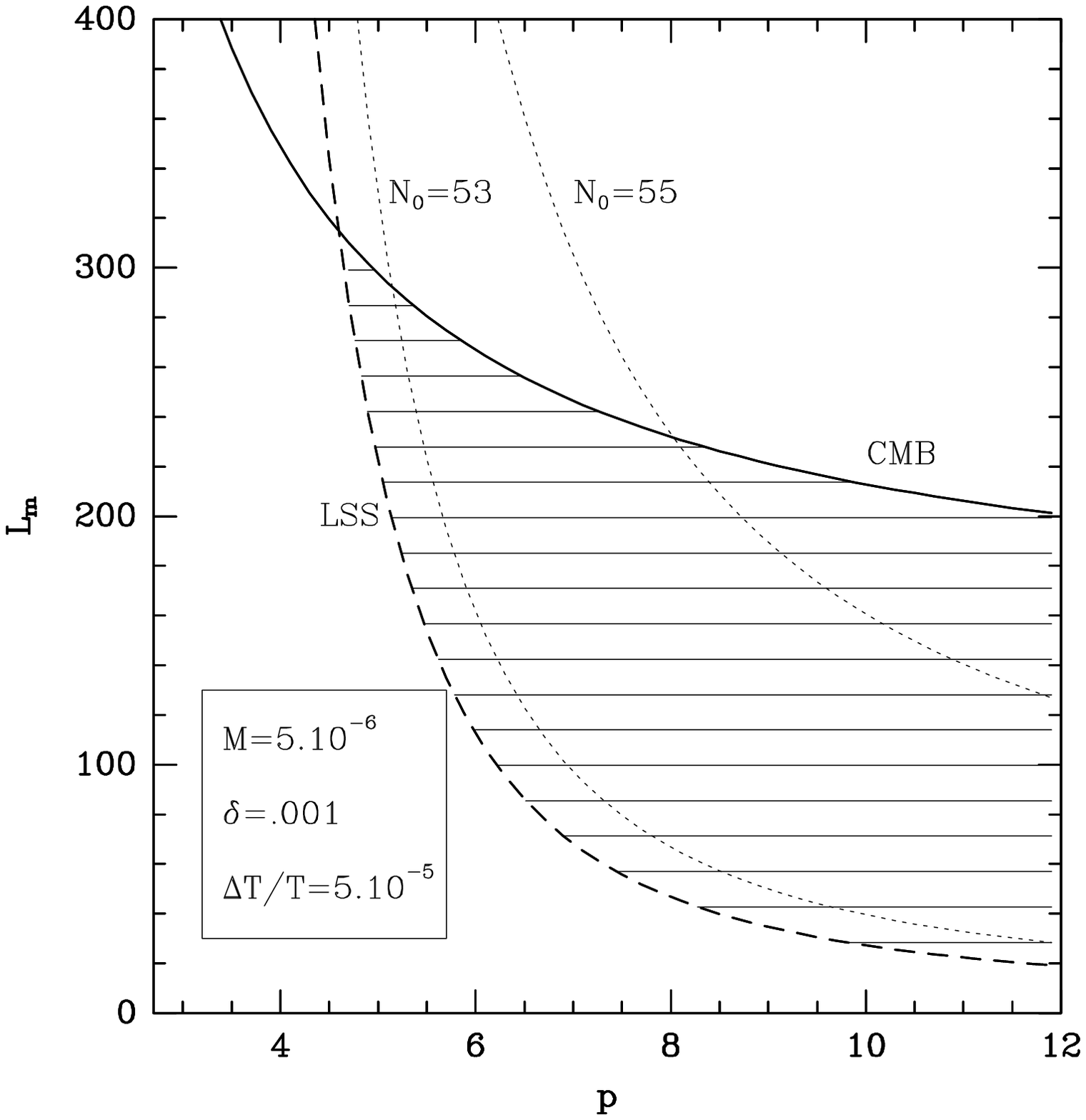}

\end{document}